\documentclass[aps,showpacs]{revtex4}
\newcommand{\cd}{\makebox[0.08cm]{$\cdot$}}
\usepackage{epsf}    
\usepackage{dcolumn}

\begin{document}

\title{Stability of bound states in the light-front Yukawa model}

\author{M. Mangin-Brinet, J. Carbonell}
\affiliation{Institut des Sciences Nucl\'{e}aires,
        53, Av. des Martyrs, 38026 Grenoble, France}%
\author{V.A. Karmanov}
\affiliation{Lebedev Physical Institute, Leninsky Pr. 53, 117924 Moscow, Russia}%

\date{\today}

\begin{abstract} We show that in the system of two fermions interacting by
scalar  exchange, the solutions for J$^{\pi}$=$0^+$ bound states are stable 
without any cutoff regularization for coupling constant below some critical
value. 
\end{abstract}


\maketitle

\section{Introduction}\label{introduc}

A promising approach to solve the bound-state problem in field theory is the
Light-Front Dynamics (LFD) \cite{perry,BPP_PR_98,Bakker,cdkm}.  In this
approach, the state vector is defined  on the surface $t+z=0$. The bound states
in the Yukawa model (two fermions interacting by scalar exchange) were studied
in  papers  \cite{glazek1,glazek2} using Tamm-Dancoff method.  It was found in
particular that because the dominating kernel  at large momenta tends to a
constant,  the binding energy of the $J=0^+$ state is cutoff dependent, what
requires the  renormalization of Hamiltonian.

The two-fermion wave functions were also considered in the explicitly
covariant version of LFD \cite{karm76}. 
The state vector is there defined on the plane given by the
invariant  equation  $\omega\cd x=0$ with 
$\omega^2=0$. The wave functions for the deuteron \cite{ckj1} and
the $pn$ scattering $J^{\pi}=0^+$ state \cite{ckj0} were found in a
perturbative way and successfully applied to calculating the
deuteron e.m. form factors \cite{ck-epj} recently measured at TJNAF \cite{t20}.
In papers \cite{ckj1,ckj0}  the $NN$ interaction was taken from the Bonn model
which includes  scalar, pseudoscalar and vector boson exchanges  with cutoffs
in the $NN$-meson vertices.  The cutoff dependence of the spectrum was not
analysed. LFD equations have now been solved exactly for a two fermion  system
in the ladder approximation with different boson exchange couplings
\cite{Mariane}.

In the present paper, we  consider the
solutions of the LFD equations for the scalar exchange only.  We
investigate the stability of the bound state solutions relative to the cutoff,
disregarding the self energy contribution and renormalization.  
We have found a critical phenomenon
for the cutoff dependence of the energy levels. The $J=0^+$ solutions are
stable -- not cutoff dependent when it tends to infinity -- for coupling constant
below some critical value. On the contrary, for coupling constants exceeding
the critical value, the system is unstable (collapses) even in  absence of the
most  singular interaction kernel in the Hamiltonian matrix. 
Our results are compared to those obtained in \cite{glazek1}.

In section \ref{equat} we write down the equations in a convenient form for
analysis. In section \ref{depend}, the cutoff dependence of the energy  levels is
studied analytically. In section \ref{num}, the results of numerical 
calculations are given. Section \ref{concl} contains the concluding remarks. 

\section{Equations}\label{equat}

We  start with the system of equations for the two spin components 
$\Phi_{1}$ and $\Phi_{2}$ which
contain a superposition of singlet and triplet spin states:
\begin{eqnarray}\label{eq4}
\left(M^2-M_0^2\right)\Phi_1(R_{\perp},x)&=&\frac{\alpha}{4\pi^2}\int 
\left[V_{11}(R_{\perp},x;R_{\perp}',x')\Phi_1(R_{\perp}',x')
+V_{12}(R_{\perp},x;R_{\perp}',x')\Phi_2(R_{\perp}',x')
\right]R_{\perp}'dR_{\perp}'dx', \nonumber\\
\left(M^2-M_0^2\right)\Phi_2(R_{\perp},x)&=&\frac{\alpha}{4\pi^2}\int  
\left[V_{21}(R_{\perp},x;R_{\perp}',x')\Phi_1(R_{\perp}',x')
+V_{22}(R_{\perp},x;R_{\perp}',x')\Phi_2(R_{\perp}',x')\right]
R_{\perp}'dR_{\perp}'dx',
\end{eqnarray}
with $\alpha={g^2\over4\pi}$, $M_0^2={R_{\perp}^2+m^2\over x(1-x)}$ 
and kernels 
$V_{ij}$ given by: 
\begin{eqnarray}\label{eqap2}	  
V_{ij}&=&\int_0^{2\pi}v_{ij}
[(K^2+\mu^2)x(1-x)x'(1-x')]^{-1} d\phi'\nonumber\\
v_{11}&=&R_\perp R_{\perp}'(2xx'-x-x')
+\left[R_{\perp}^2 x'(1-x')+ R'^2_{\perp} x(1-x)-m^2(x+x')(2-x-x')\right]
\cos\phi',\nonumber\\
v_{12}&=&m \left[R_\perp (x+3x'-2xx'-2x'^2)
-R'_\perp (x'+3x-2x'x-2x^2)\cos\phi'\right],\nonumber\\
v_{21}&=&m \left[R'_\perp (x'+3x-2x'x-2x^2)
-R_\perp (x+3x'-2xx'-2x'^2)\cos\phi'\right],\nonumber\\
v_{22}&=&R_\perp R_{\perp}'(2xx'-x-x')\cos\phi'
+\left[R_{\perp}^2 x'(1-x')+ R'^2_{\perp} x(1-x)-m^2(x+x')(2-x-x')\right].
\end{eqnarray}
The momentum $K^2$ reads (see eq. (7.27) from \cite{cdkm}):
\begin{equation}\label{kernrx}
K^2= m^2{x'\over x}\left(1-\frac{x}{x'}\right)^2
+\frac{x'}{x}R_{\perp}^2-2R_{\perp}R_{\perp}'\cos\phi'+\frac{x}{x'}
R_{\perp}'^2+(x'-x)\left(\frac{m^2+R_{\perp}'^2}{x'(1-x')}-M^2\right)
\end{equation}
for $x\leq x'$ and with the replacements $x\leftrightarrow x'$ 
$R_{\perp}\leftrightarrow R_{\perp}'$ for $x\geq x'$.
Functions $\Phi_{1,2}$  are normalized as:
\begin{equation}\label{eq3a}
\int\left\{|\Phi_1(R_{\perp},x)|^2 +|\Phi_2(R_{\perp},x)|^2\right\} 
R_{\perp}dR_{\perp}dx =1.
\end{equation}
Equations (\ref{eq4}) and kernels (\ref{eqap2}) are taken from \cite{glazek1}
(eqs. 3.1a-b and C1-C4) with the notations 
$\Phi^{1+}\equiv \Phi_1,\Phi^{2-}\equiv \Phi_2$ and
$k\equiv R_{\perp}$, $q\equiv R'_{\perp}$, $y\equiv x'$.

In view of further analysis it is useful to introduce the new functions 
$f_{1,2}$ given by:
$$\Phi_i={1\over N(x)}\sum_j c_{ij}(R_\perp) f_j $$
$$
\mbox{with}\qquad\qquad
N(x)={2^{3/2}\pi\over\sqrt{m}}\sqrt{x(1-x)}, \qquad
\hat{c}={1\over\sqrt{R_\perp^2+m^2}}\pmatrix{-R_\perp& m\cr m &R_\perp}$$
and the variables $k,\theta$ ($k\geq 0$, $0\leq\theta\leq\pi$) defined by:
\begin{equation}\label{var}
R_{\perp}=k\sin\theta,\quad
x=\frac{1}{2}\left(1-\frac{k\cos\theta}{\sqrt{k^2+m^2}}\right).
\end{equation}
In this new representation, the normalization condition (\ref{eq3a}) obtains
the form:
\begin{equation}\label{na5} 
{m\over (2\pi)^3}\int \left\{f_1^2(k,\theta)+ f_2^2(k,\theta)\right\} 
{d^3k\over\varepsilon_k}=1,
\end{equation}
where $\varepsilon_k=\sqrt{m^2+k^2}$,
and one gets for $f_{i}$ the system of equations:
\begin{eqnarray}\label{eq10a}  
\left[4(k^2 +m^2)-M^2\right] f_1(k,\theta) &=&-\frac{m^2}{2\pi^3} \int
\left[K_{11}(k,\theta;k',\theta')
f_1(k',\theta')+K_{12}(k,\theta;k',\theta')
f_2(k',\theta')\right]\frac{d^3k'}{\varepsilon_{k'}},\nonumber\\
\left[4(k^2 + m^2)-M^2\right] f_2(k,\theta) &=&
-\frac{m^2}{2\pi^3} \int \left[K_{21}(k,\theta;k',\theta')
f_1(k',\theta')+K_{22}(k,\theta;k',\theta')
f_2(k',\theta')\right]\frac{d^3k'}{\varepsilon_{k'}}.
\end{eqnarray}
For convenience, we keep in (\ref{eq10a}) the integration over $d\phi'$,
though it has been already performed in $K_{ij}$.  
The kernels $K_{ij}$ are linearly expressed in terms of $V_{ij}$
and their analytical expressions read: 
\begin{eqnarray}\label{eqap1}
K_{ij}&=&\int_0^{2\pi}{\kappa_{ij} \over
(K^2+\mu^2)m^2\varepsilon_k \varepsilon_{k'}}{d\phi'\over 2\pi}, \nonumber\\
\kappa_{11}&=&-\alpha\pi
\left[2 k^2 k'^2+3k^2 m^2+3k'^2 m^2+4 m^4
-2 k k'\varepsilon_k \varepsilon_{k'} \cos\theta \cos\theta' 
- k k' (k^2 + k'^2 + 2 m^2)
\sin\theta \sin\theta' \cos\phi'\right],\nonumber\\
\kappa_{12}&=&-\alpha\pi m
(k^2 - k'^2) \left(k'\sin\theta' + k\sin\theta\cos\phi' \right),
\nonumber\\
\kappa_{21}&=&-\alpha\pi m
(k'^2 - k^2) \left(k\sin\theta + k'\sin\theta'\cos\phi' \right),
\\
\kappa_{22}&=&-\alpha\pi
\left[\left(2 k^2 k'^2+3k^2 m^2+3k'^2 m^2+4 m^4
- 2 k k' \varepsilon_k\varepsilon_{k'}
\cos\theta \cos\theta'\right)\cos\phi'
-k k'(k^2 + k'^2 + 2 m^2) \sin\theta\sin\theta'\right],	
\nonumber
\end{eqnarray}
and (see eq. (3.60) from \cite{cdkm}):	
\begin{eqnarray*}
K^2 &=& k^2+k'^2
-2kk'\left(1+\frac{(\varepsilon_k -\varepsilon_{k'})^2}
{2\varepsilon_k\varepsilon_{k'}}\right)\cos\theta\cos\theta'-2k k'\sin\theta \sin\theta'\cos\phi'
+\left(\varepsilon_{k}^2
+\varepsilon_{k'}^2-\frac{1}{2}M^2\right)                           
\left|\frac{k\cos\theta}{\varepsilon_{k}}                                
-\frac{k'\cos\theta'}{\varepsilon_{k'}}\right|.                               
\end{eqnarray*}
In the variables $k,\theta$, the kinetic energy in (\ref{eq10a}) is  quadratic
on $k$, the kernels are smooth in $\theta$, and the stability of the binding
energy is related to the kernels behavior at large $k$. 

In the explicitly covariant version of LFD, the states are labeled by the
eigenvalues $J$ corresponding to the appropriate angular momentum  operator
\cite{cdkm} $\vec{J}=\vec{J}_0+\vec{J}_{\vec{n}}$, containing, besides the free
operator $\vec{J}_0$,  the term $\vec{J}_{\vec{n}}=-i[\vec{n}\times\partial
\vec{n}]$, where in c.m.-system $\vec{n}=\vec{\omega}/|\vec{\omega}|$.
The two fermion wave function with $J=0$ is determined by two
spin components \cite{ckj0}. The $J=1$ wave function is determined by
six  components \cite{ckj1} and the equations are split in two
subsystems, distinguished by the eigenvalues $a=0,1$ of 
$\hat{A}=(\vec{J}\cd \vec{n})^2$.
The $J=1,a=0$ subsystem includes two components and
the $J=1,a=1$ one includes four. The coupled equations 
obtained in
this way (two for $J=0$ and two+four for $J=1$) correspond to the
2+2+4 systems from \cite{glazek1}. In this classification, the equations 
(\ref{eq10a}) with the kernels (\ref{eqap1}) are written for the $J=0$ state,
corresponding to $(1+,2-)$ from \cite{glazek1}. In the explicitly covariant LFD
these equations are directly obtained in the form (\ref{eq10a}). 
Their direct derivation, together with the $J=1$ case,
will be given in a more detailed publication \cite{Mariane}.
The relation
between $\Phi_i$ and $f_i$  and the corresponding relation between  $K_{ij}$
and $V_{ij}$ result from  the different representations of the spinors used in 
\cite{glazek1} and \cite{cdkm}.  Below we present also  the results  for the
$J=1,a=0$ state, whose components are related by a linear combination to
$\Phi^{1-},\Phi^{2+}$ defined in \cite{glazek1}. To distinguish the states,
when necessary, we will attach to the kernels the index $J=0$ or $J=1$ and omit
$a=0$.   

We would like to emphasize that the equations (\ref{eq4}) solved in
\cite{glazek1} are related to our explicitly covariant LFD equations
(\ref{eq10a}) only by a linear tranformation of the components and a variable
change. Both equations are thus strictly equivalent.

\section{The cutoff dependence of the binding energy}\label{depend}

We consider the equations on the finite interval $0<k<k_{max}$.
The dependence of the solution on the cutoff $k_{max}$ in the limit $k_{max}\to
\infty$ is determined by the kernels asymptotics. Let us first analyze
the kernel  $K_{11}$. In the $(k,k')$-plane, when both $k,k'\to\infty$ 
with a fixed ratio $k'/k=\gamma$, this kernel tends to a constant. From the
expressions (\ref{eqap1}) we find the asymptotics: 
\begin{equation}\label{eqn16}
K_{11}=-\frac{2\pi^2}{m}\left\{
\begin{array}{ll}
\sqrt{\gamma}A_{11}(\theta,\theta',\gamma),
& \mbox{if $\gamma<1$}
\\
\frac{\displaystyle{A_{11}(\theta,\theta',1/\gamma)}}
{\displaystyle{\sqrt{\gamma}}},
& \mbox{if $\gamma>1$}
\end{array}\right.
\end{equation}
with
\begin{equation}\label{eq14a}
A_{11}(\theta,\theta',\gamma)=
\frac{\alpha'}{\sqrt{\gamma}}\int_0^{2\pi}\frac{d\phi}{2\pi}
\frac{2\gamma(1-\cos\theta\cos\theta')-
(1+\gamma^2)\sin\theta\sin\theta'\cos\phi}
{(1+\gamma^2)(1+|\cos\theta-cos\theta'|-\cos\theta\cos\theta')
-2\gamma\sin\theta\sin\theta'\cos\phi},
\end{equation}
and $\alpha'=\alpha/(2m\pi)$.  For convenience, we extracted the factor 
$\sqrt{\gamma}$ in eq. (\ref{eqn16}). When $k'$ is fixed and $k\to \infty$, 
$K_{11}$ decreases like $1/k$, and analogously for $k$ fixed, $k'\to \infty$. 

We will compare the kernels in eq. (\ref{eq10a}) with the kernel corresponding
to the non-relativistic potential $U(r)=-\alpha'/r^2$.  To get the exact
correspondence, we take the S-wave  Schr\"odinger equation   for the wave
function $\psi(k)$, with the kernel $\tilde{U}(k,k')$, Fourier transform of
$U(r)=-\alpha'/r^2$, and we
substitute there $\psi(k)=f(k)\sqrt{m/\varepsilon_{k}}$. The equation obtained
for $f(k)$ has the relativistic form (\ref{eq10a}). It contains the factor
$1/\varepsilon_{k'}$ in the integration volume and the kernel
$K_{-\alpha'/r^2}(k,k')
=\tilde{U}(k,k')\sqrt{\varepsilon_{k}\varepsilon_{k'}}/m$.  For
$k'/k=\gamma$ fixed the latter has the asymptotics (\ref{eqn16}) with the
constant  $\alpha'$ instead of the function $A_{11}$.  For $k'$ fixed, $k\to
\infty$ it decreases like $1/\sqrt{k}$, and analogously for $k$ fixed,  $k'\to
\infty$.

The kernel $K_{22}$, when $k\to \infty$ and $k'/k=\gamma$ fixed, has also the
asymptotics (\ref{eqn16}), however with an unbounded function $A_{22}$, in
contrast to $A_{11}$ (see below). When $k\to \infty$ and $k'$ fixed (and vice
versa) it tends to a constant. Therefore it always dominates over  $K_{11}$ and
over $K_{-\alpha'/r^2}(k,k')$.

To disentangle the two different sources of collapse, we first consider the one
channel problem  for the component $f_1$ with the kernel $K_{11}$,  and we
remove the second equation from (\ref{eq10a}). We analyse the domain of
$k'/k=\gamma$ fixed, where $K_{11}$ has its
maximal asymptotics values. The $\theta,\theta'$ domain is finite
and the
function  $A_{11}(\theta,\theta',\gamma)$ has no singularities
in $\theta,\theta'$. Therefore we
majorate it by its maximal value. If the stronger, majorated kernel results in
stable bound states, the exact one results in stability too.  This method to
analyse the cutoff dependence is equivalent to applying the sufficient
condition of stability proposed in \cite{ss99}. The
inspection of (\ref{eq14a}) shows that for fixed $\gamma$, the maximum of
$A_{11}$ is achieved at $\theta=\theta'$, where it has
the form: 
$
A_{11}(\theta=\theta',\gamma)=\alpha'\sqrt{\gamma}
$
independent of $\theta$.
Note that 
$
A_{22}(\theta=\theta',\gamma)=\alpha'/\sqrt{\gamma}
$
is unbounded when $\gamma\to 0$. Then we majorate the function
$A_{11}(\theta=\theta',\gamma)$ by its maximal value relative to $\gamma$: 
$A^{max}_{11}=\alpha'$, which is evidently achieved at $\gamma=1$. With this
value of $A_{11}$, the kernel (\ref{eqn16}) exactly  coincides with the kernel
$K_{-\alpha'/r^2}(k,k')$. As well known \cite{ll}, with the potential 
$-\alpha'/r^2$, the binding energy does not depend on cutoff if
$\alpha'<1/(4m)$, what restricts the coupling constant to
$\alpha<\pi/2$. If $\alpha'>1/(4m)$, the system collapses, what is manifested
by the fact that
the binding  energy tends to $-\infty$ when $k_{max}\to \infty$.  
In this system, there exists
a critical value $\alpha_{c}$ of the coupling constant, 
below which the binding energy is stable.
Majorating the kernel, we underestimate $\alpha_{c}$. We
calculated also this value \cite{Mariane}, not majorating the function
$A_{11}=\alpha'\sqrt{\gamma}$, but taking into account its dependence on
$\gamma$. In this way we find  $\alpha_{c}=\pi$, instead of $\pi/2$, when 
$\sqrt{\gamma}$ was replaced by 1.  Because of majorating the kernel in the
variables $\theta,\theta'$, this value  should be still smaller than the true
$\alpha_{c}$ but it is in the range $3<\alpha_c<4$ we found numerically (see
next section).

In the two-channel problem, the kernel  dominating in  asymptotics is $K_{22}$.
In the case $J=0$ it is positive and corresponds to repulsion. Because of that,
this channel does not lead to any collapse. This repulsion cannot prevent from
the collapse in the first channel (for enough large $\alpha$), since due to
coupling between two  channels the singular potential in the channel 1 "pumps
out" the wave function from the channel 2 into the channel 1. So, in the
coupled equations system  (\ref{eq10a}) the situation with the cutoff
dependence is the same as for one channel.

Let us now consider the state $J=1,a=0$. The asymptotics of  the kernel
$K^{(J=1)}_{22}$  is the same than $-K^{(J=0)}_{22}$, it is negative and
corresponds to attraction.  Since it always dominates over
$K_{-\alpha'/r^2}(k,k')$, this attraction is stronger than in the
$-\alpha'/r^2$ potential. Therefore it results in a collapse for any value of the
coupling constant. In the paper \cite{glazek1}, this situation corresponds to
the two-fermion state described by the components $\Phi^{1-},\Phi^{2+}$.

\section{Numerical results}\label{num}

\begin{figure}[hbtp]
\begin{minipage}[t]{77mm}
\begin{center}
\mbox{\epsfxsize=7.5cm\epsfysize=6.0cm\epsffile{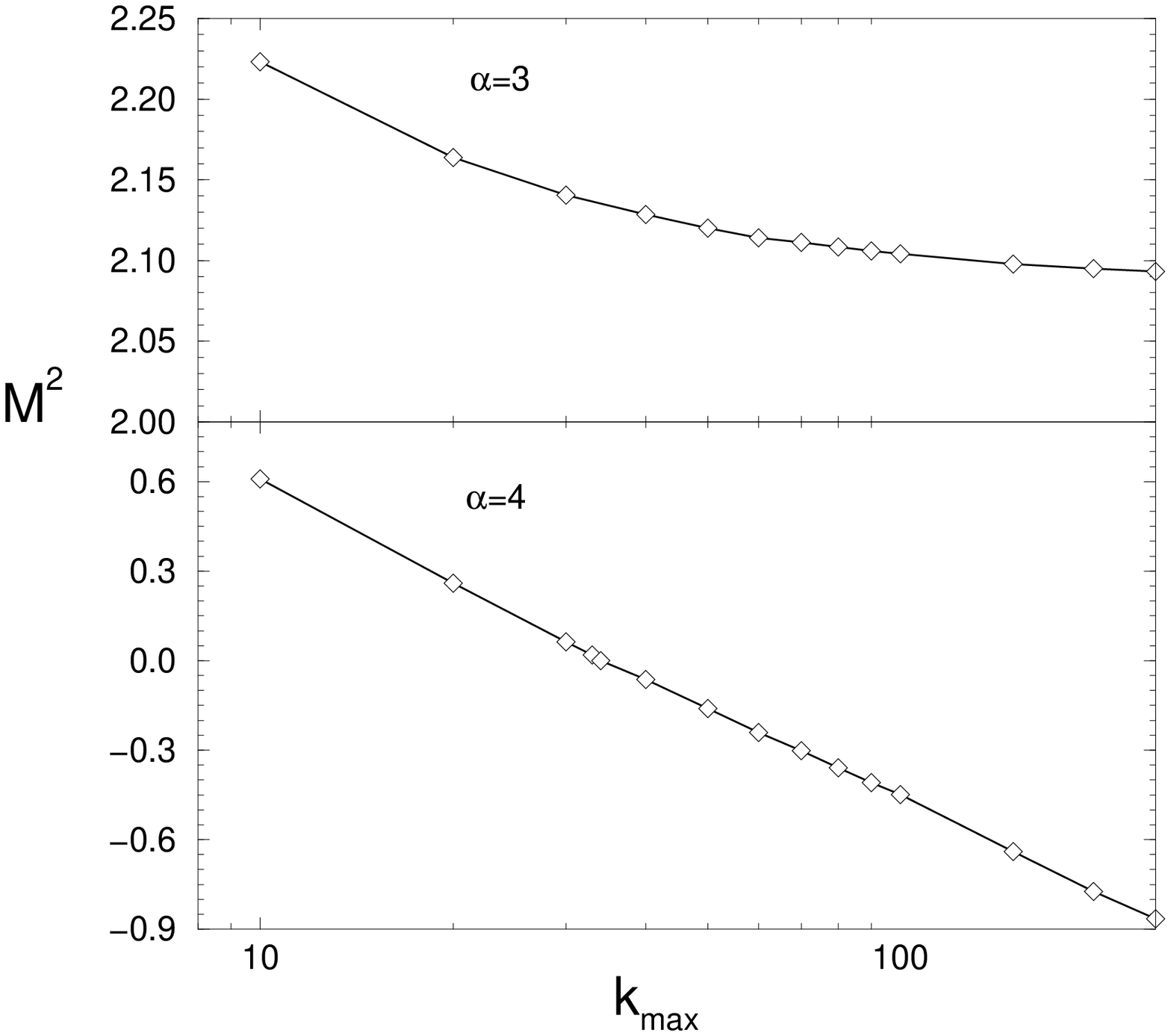}}
\end{center}
\caption{Cutoff dependence of the binding energy in the $J=0$ or $(1+,2-)$
state,
in the one-channel problem ($f_1$), for two fixed values of the coupling 
constant below and above the critical value.}\label{B_kmax}
\end{minipage}
\hspace{0.5cm}
\begin{minipage}[t]{77mm}
\begin{center}
\mbox{\epsfxsize=7.5cm\epsfysize=7.5cm\epsffile{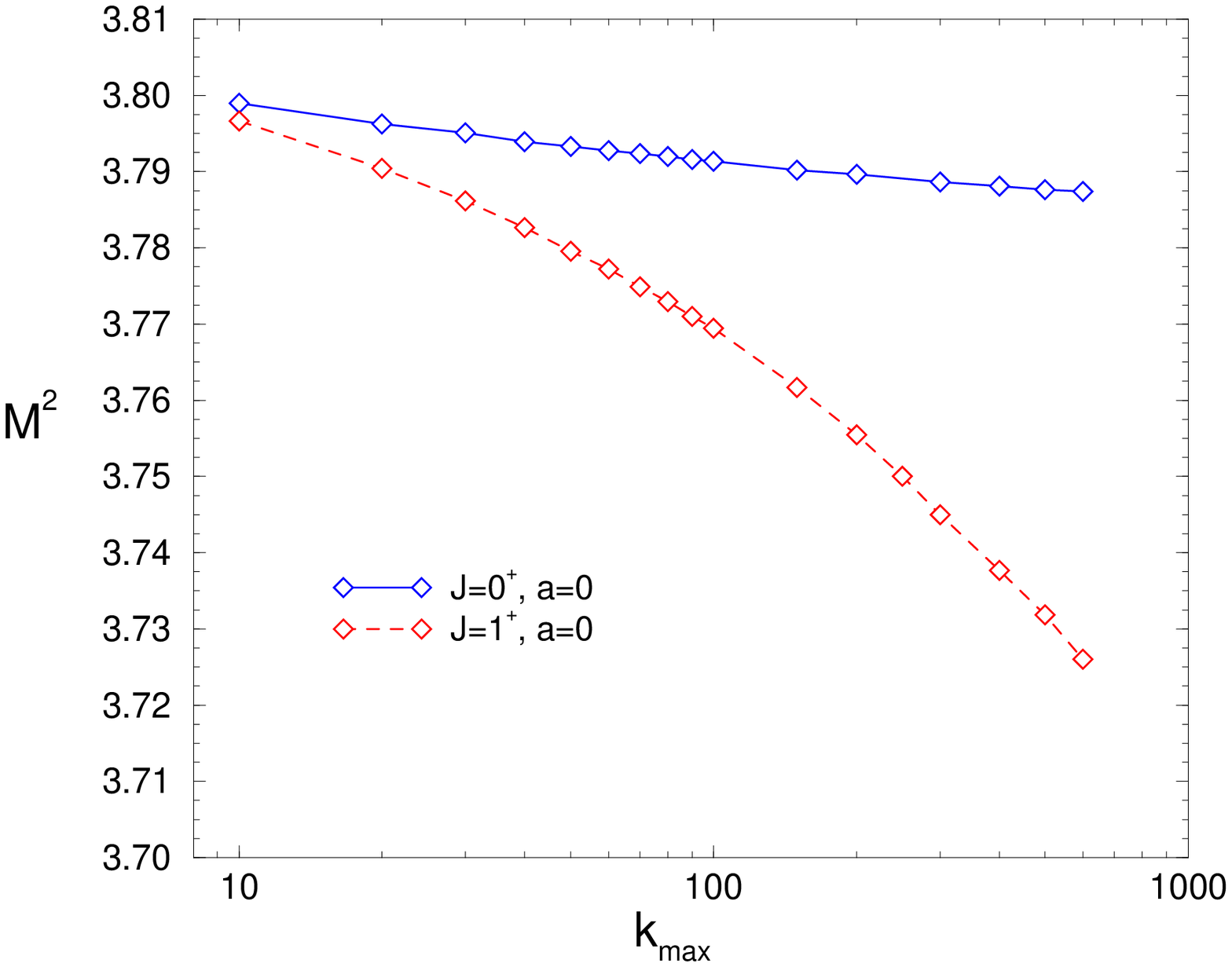}}
\end{center}
\caption{Cutoff dependence of the binding energy, for $J=0$  $(1+,2-)$ and 
$J=1,a=0$ $(1-,2+)$ states, in the two-channel problem  
($\alpha=1.184$).}
\label{alpha_kmax_500}
\end{minipage}
\end{figure}

The preceding results are confirmed by numerical calculations.
The constituent masses were taken equal to $m$=1 and the mass of the exchanged
scalar $\mu$=0.25.

We first present the results given by the single equation for $f_1$ with 
kernel $K_{11}$ in the $J=0$ case. We have plotted in figure \ref{B_kmax} the
mass square $M^2$ of the two fermion system  as a function of the cutoff
$k_{max}$ for two fixed values of the coupling constant below and above
the critical value $\alpha_{c}$. 
In our calculations,  the cutoff appears directly as the maximum
value $k_{max}$  up to which the integrals in (\ref{eq10a}) are performed. One
can see two dramatically different behaviors depending on the value of the
coupling constant $\alpha$. For $\alpha=3$, i.e.  $\alpha<\alpha_{c}$,  the
result is convergent. For $\alpha=4$, i.e. $\alpha>\alpha_{c}$, the result is
clearly divergent. $M^2$ decreases logarithmically as a function of $k_{max}$
and becomes even negative. We would like to notice that this divergence is not
associated with the non decreasing behavior of the $K_{22}$ kernel but with the
existence of a critical value of the coupling constant  separating two
dynamical regimes. This property is due only to the large $k$ behavior of
$K_{11}$.  Though the negative values of $M^2$ are physically meaningless, they
are formally  allowed by the equations (\ref{eq4}) and (\ref{eq10a}). The first
degree of $M$ does not enter neither in the equation nor in the kernel, and
$M^2$  crosses zero without any singularity.  The value of
$\alpha_c$ does not depend on the exchange mass $\mu$. For $\mu\ll m$, e.g.
$\mu\approx0.25$, its existence is not relevant in describing physical states  
since any solution with positive $M^2$, stable relative to cutoff, corresponds
to $\alpha<\alpha_c$. For $\mu\sim m$ one can reach the critical $\alpha$  for
positive, though small values of $M^2$.

We consider now the full Yukawa problem as given by  the two coupled equations
(\ref{eq10a}).  In figure \ref{alpha_kmax_500} are displayed the variations of
$M^2$ for $J=0$ or $(1+,2-)$ and $J=1,a=0$ or $(1-,2+)$ states as a function
of the cutoff $k_{max}$. The value of the coupling constant for both $J$  is
$\alpha=1.184$, the same that in Fig. 2 of \cite{glazek1}, below the critical
value. Our numerical values are in agreement with the results  for the cutoff
$\Lambda \leq 100$  presented in this figure \cite{glazek1}, but our
calculation at larger $k_{max}$ leads to different conclusion for the $J=0$
state. We first notice a qualitatively different behavior of the two states. In
what concerns  $J=0$, the curve becomes flat when $k_{max}$
increases,  -- with a 0.1\% variation in $M^2$ when changing
$k_{max}$ from 50 to 600. 
We thus conclude to the stability of the state with $J=0$, as expected from our
analysis in sect. \ref{depend}.

On the contrary, for $J=1,a=0$ the value of  $M^2(k_{max})$
decreases  faster than logarithmically what indicates -- as found in
\cite{glazek1} -- a collapse. As mentioned above, the asymptotics of the 
$K^{(J=1)}_{22}$ kernel is the same as the $K^{(J=0)}_{22}$ one but with an
opposite sign, i.e. it is attractive, what leads to unstability for any value
of $\alpha$. We found the same result when solving the $J=0$ equations with
the opposite sign of $K^{(J=0)}_{22}$.

\section{Conclusion}\label{concl}

The Light-Front solutions of the two fermion system interacting via a scalar
exchange have been obtained. We have found that the $J=0$ -- or
$(1+,2-)$ -- state is stable (i.e. convergent relative  to the cutoff $k_{max}\to
\infty$) for  coupling constant below some critical value, in a way similar
to what is known in non relativistic quantum mechanics for the $-\alpha'/r^2$
potential. In this point, our conclusion differs from the one settled in
\cite{glazek1}, where it was stated that the integrals in eqs. (\ref{eq4})
diverge  logarithmically with cutoff.  Above the critical value the system
collapses, what manifests as an unbounded cutoff dependence of $M^2$. In the
$J=1,a=0$ -- or $(1-,2+)$ -- state the system is found to be always unstable, in
agreement with  \cite{glazek1}.

The origin of this unstability for the $J=0$ state  differs from $J=1$.  We
have found that the $K_{22}^{(J=0)}$ dominating kernel does not generate 
a collapse because it is repulsive. 
The unstability in this case is related to $K_{11}$.

These results should be taken into account when carrying out
the renormalization procedure. 
The explicitly covariant LFD may be efficient for solving this problem,
like it has proved to be fruitful for analyzing the Yukawa model.

\bigskip
{\bf Acknowledgements:}
The authors are indebted to T. Frederico for pointing out the reference
\cite{glazek1} and to F. Coester, St. Glazek and A.V. Smirnov for useful
discussions. One of the authors (V.A.K.) is sincerely grateful for the warm
hospitality of the theory group at the Institut des Sciences Nucl\'{e}aires,
Universit\'e Joseph Fourier, in Grenoble, where this work was performed. 
The numerical calculations were performed   at CGCV (CEA Grenoble) and  IDRIS
(CNRS). We thank the staff members  of these two organizations for their
constant support.

\end{document}